\documentclass[conference,a4]{IEEEtran}

\usepackage[mathcal]{euscript}
\usepackage{epsfig}
\usepackage{mathrsfs}
\usepackage{stmaryrd}
\usepackage{amsmath,amstext,amsfonts,amssymb}
\usepackage{epsfig,float}
\usepackage{graphicx}
\usepackage{cite}
\usepackage{psfrag}
\usepackage{subfigure}
\usepackage{url}
\usepackage{shadow,color,pifont,times,rotate}
\DeclareMathAlphabet{\mathpzc}{OT1}{pzc}{m}{it}

\newcommand{\mb}{\mathbf}

\newcommand{\mc}{\mathcal}

\newtheorem{theorem}{Theorem}[section]

\renewcommand{\baselinestretch}{1.2}
\def\Mcb{M}

\begin{document}


\title{Dirty-paper Coding without Channel Information\\ [-3mm] at the Transmitter and Imperfect \\ [-3mm] Estimation at the Receiver\vspace*{-2mm}}


\author{Pablo Piantanida and Pierre Duhamel\\
Laboratoire des Signaux et Syst\`emes, CNRS/Sup\'{e}lec,\\
 F-91192 Gif-sur-Yvette, France \\
Email:\{piantanida,pierre.duhamel\}@lss.supelec.fr }

\maketitle

\begin{abstract}
In this paper, we examine the effects of imperfect channel estimation at the receiver and no channel knowledge at the transmitter on the capacity of the fading Costa's channel with channel state information non-causally known at the transmitter. We derive the optimal Dirty-paper coding (DPC) scheme and its corresponding achievable rates with the assumption of Gaussian inputs. Our results, for uncorrelated Rayleigh fading, provide intuitive insights on the impact of the channel estimate and the channel characteristics (e.g. SNR, fading process, channel training) on the achievable rates. These are useful in practical scenarios of multiuser wireless communications (e.g. Broadcast Channels) and information embedding applications (e.g. robust watermarking). We also studied optimal training design adapted to each application. We provide numerical results for a single-user fading Costa's channel with maximum-likehood (ML) channel estimation. These illustrate an interesting practical
  trade-off between the amount of training and its impact to the interference cancellation performance using DPC scheme.


\end{abstract}

\IEEEpeerreviewmaketitle

\section{Introduction}

Consider the problem of communicating over a Gaussian channel corrupted by an additive Gaussian interfering signal that is non-causally known at the transmitter. This variation of the conventional additive white Gaussian noise (AWGN) channel is commonly known as {\it channel with state information at the transmitter}. The state $S$ is a random Gaussian variable with power $Q$ and independent of the Gaussian noise $Z$. The channel input is the message $m \in \{1,\dots,\lfloor2^{nR}\rfloor \}$ and its output is $Y=X+S+Z$, where $R$ is the rate in bit per transmission. The capacity expression of single-user channels with random parameters has been derived by Gel'fand and Pinsker in \cite{pinsker-1980}. The authors show that the capacity of such a channel $\{W(y|x,s),x\in\mc{X},s\in\mc{S}\}$ with state information $S$ non-causally available at the transmitter is\vspace{-2mm}
 \begin{equation}
C=\sup\limits_{p(u,x|s)}\big\{I(U;Y)-I(U;S)\big\},\vspace{-2mm}
\end{equation}
$U$ is an auxiliary random variable chosen so that $U\minuso (X,S) \minuso Y$ form a Markov Chain and $p(u,x|s) = \delta\big(x - f(u,s)\big)p(u|s)$.

In ``Writing on Dirty Paper'' \cite{costa-1983}, Costa applied this result to an AWGN channel corrupted by an additive white Gaussian interfering signal $S$. He showed that choosing $U = X + \alpha S$, with an appropriate value for $\alpha$ ($\alpha^* =\bar{ P}/(\bar{P}+\sigma_Z^2)$, $\sigma_Z^2$ being the AWGN variance). This coding scheme, referred as \emph{Dirty-paper coding (DPC)}, allows one to achieve the same capacity as if the interfering signal $S$ was not present, i.e. $C=\frac{1}{2}\log_2 \left(1+\frac{\bar{P}}{\sigma_Z^2}\right)$. This result has gained considerable attention during the last years, mainly because of its potential use in communication scenarios where interference cancellation at the transmitter is needed. In particular, multiuser interference cancellation  for Broadcast Channels (BC) and information embedding (digital watermarking for multimedia security applications) are instances  of such scenarios. In the recent years, the Gaussian Multiple-Input-Multiple-Output Broadcast Channel (MIMO-BC) has been extensively studied. In \cite{caire-2003}, the authors based on DPC have established an achievable rate region, referred to as \emph{Dirty-paper coding region}. Recently in \cite{steinberg-shamai-MIMO-BC},  the DPC region was proved to be equal to the capacity. 

Most of the literature focuses on the information-theoretic performances of DPC under the assumption on the availability of perfect channel information at both transmitter and receiver. However, it is well-known that the performances of wireless systems are severely affected if only a noisy estimate that differs from the true channel is available (cf. \cite{medard-2000}, \cite{goldsmith_04} and \cite{spawc06}). Of particular interest is the issue of the effect of this imperfect channel knowledge if interference cancellation or Dirty-paper coding is used. The problem may even be more serious  in the practical situations where no channel information is available at the transmitter, i.e., no feedback information from the receiver back to the transmitter with the channel estimates. 

Throughout this paper, we consider a wireless or watermarked channel modeled as $Y=H(X+S)+Z$, where $H$ is the random channel, which neither the transmitter nor the receiver know. We assume that the receiver estimates $H$ during a phase of independent training, by using maximum-likelihood (ML) channel estimation (Section III). Whereas, the transmitter does not know this estimate. Then, we observe that depending on the targeted application, e.g. Broadcast Channel or robust watermarking, two different training scenarios are relevant. In this work, we determine the tradeoff between the amount of training required for channel estimation and the corresponding achievable rates using DPC (Section IV). We address this problem through the notion of reliable communication based on the average of the error probability over all channel estimation errors. This allows to make an equivalence with the capacity of a composite (more noisy) channel. Our proposed framework is sufficiently general 
 to involve the most important information embedding and multiuser communication scenarios. Finally, Section V uses a Rayleigh-fading Costa's channel to illustrate average rates over all estimates, for different amount of training. 


\section{Channel model}

First consider a general model for communication under channel uncertainty over discrete memoryless channels (DMCs) with input alphabet $\mathscr{X}$, output alphabet $\mathscr{Y}$ and channel states $\mathscr{S}$ (cf. \cite{pinsker-1980} and \cite{lapidoth-resume}).  A specific instance of the unknown channel is characterized by a transition probability mass (PM) $W(\cdot|x,s,\theta)\in\mc{W}_{\Theta}$ with a random state $s\in \mathscr{S}$ perfect known by the transmitter and a fixed but unknown channel $\theta\in \Theta\subseteq \mathbb{C}^d$. Here, $\mc{W}_{\Theta}=\big\{W(\cdot|x,s,\theta)\!: \, x\in \mathscr{X}, s\in \mathscr{S},\,\theta\in \Theta\big\}$ is a family of conditional transition PMs on $\mathscr{Y}$, parameterized by a vector $\theta\in \Theta$, which follows i.i.d. ~$\theta\sim\psi(\theta)$. It is assumed that the receiver only knows an estimate $\hat{\theta}$ of the channel and a characterization of the estimator performance in terms of the conditional probability density function (pdf) $\psi(\theta|\hat{\theta})$ (this can be obtained using $\mc{W}_{\Theta}$ and the a priori distribution of $\theta$). On another side, the transmitter does not know the estimate $\hat{\theta}$, it only knows its statistic $\psi(\hat{\theta})$. The extension of the DMC $W(\cdot|x,s,\theta)$ to $n$ channel uses within a block is given by $W^n(\mb{y}|\mb{x},\mb{s}, {\theta})=\prod_{i=1}^{n}  W(y_i|x_i,s_i,\theta) $ where $\mb{x}=(x_1,\dots,x_n)$, $\mb{s}=(s_1,\dots,s_n)$ and $s_i$ is an i.i.d. realization of $P_S(s)$ and $\mb{y}=(y_1,\dots,y_n)$. It is assumed that the state sequence $\mb{s}$ is perfectly known at the transmitter before sending $\mb{x}$ and unknown at the receiver.

Throughout this paper we consider a memoryless fading Costa channel. The discrete-time channel at time $t$ is
\begin{equation}
Y(t)=H(t)\big(X(t)+S(t)\big)+Z(t), \label{channel-definition}
\end{equation}
where $X(t) \in \mathbb{C}$ is the transmitter symbol and $Y(t)\in \mathbb{C}$ is the received symbol. Here, $H(t)\in \mathbb{C}$ is the complex random channel ($\theta=H$) whose entries are independent identically distributed (i.i.d.) zero-mean circularly symmetric complex Gaussian (ZMCSCG) random variables $C\mc{N}(0,\sigma^2_h)$. The noise $Z(t)\in \mathbb{C}$ consists of i.i.d. ZMCSCG random variables with variance $\sigma_{Z}^2$. The channel state $S(t)\in \mathbb{C}$ consists of i.i.d. ZMCSCG random variables with variance $Q$. The quantities $H(t)$, $Z(t)$, $S(t)$ are assumed ergodic and stationary random processes, and the channel matrix $H(t)$ is independent of $S(t)$, $X(t)$ and $Z(t)$. This leads to a stationary and discreet-time memoryless channel $W\big(y|x,s,H\big)$ with pdf 
\begin{equation}
W(y|x,s,H)=C\mc{N}\big(H(x+s),\sigma_{Z}^2\big). \label{channel-model}
\end{equation}
The average symbol energy at the transmitter is constrained to satisfy $\mathbb{E}_X\{X(t)X(t)^\dag\}\leq \bar{P}$ and $(\cdot)^{\dag}$ denotes Hermitian transposition. In practical situations, only a noisy estimate $\hat{\theta}=\widehat{H}$ that differs from the true channel is available at the receiver. We next focus on training sequence design for channel estimation.

\section{Optimal design of channel training}

A standard technique to allow the receiver to estimate the channel matrix consists of transmitting training sequences, i.e., a set of symbols whose location and values are known to the receiver. We assume that the channel is constant during the transmission of an entire codeword so that the transmitter, before sending the data $\mb{x}$, sends a training sequence of $N$ symbols $\mb{x}_T=(x_{T,1},\dots,x_{T,N})$.  The average energy per training symbol is $P_T=\frac{1}{N}tr\big(\mb{x}_T \mb{x}_T^\dag\big)$. Thus, two different scenarios are relevant: 

(i) \emph{The channel affects the training sequence only}, i.e. the decoder observes $\mb{y}_{T}=H \mb{x}_T+\mb{z}_{T},$ where $\mb{z}_{T}$ is the noise affecting the transmission of training symbols. This scenario arises, e.g., in Broadcast Channels where the transmitter does not send the sequence $\mb{s}_T$ during the training phase. In that case, an optimal training is obtained by sending an arbitrary constant symbol, $x_{T,i}=x_{0}$ for all $i=1,\dots,N$. So that a maximum-likehood (ML) estimate $\hat{\theta}=\hat{H}_{\textrm{ML}}$ is obtained at the decoder from the observed output. The ML estimate of $H$ is given \cite{spawc06} by
\begin{eqnarray}
\widehat{H}_{\textrm{ML}}=\big(\mb{x}_T^\dag \mb{x}_T\big)^{-1} \mb{x}_T^\dag   \mb{y}_{T}= H+\mc{E}, \label{eq-error-1}
\end{eqnarray}
where $\mc{E} = \big(\mb{x}_T^\dag \mb{x}_T\big)^{-1} \mb{x}_T^\dag \mb{z}_{T} $ is the estimation error with
\begin{eqnarray}
\sigma_{\mc{E}}^2=\textrm{SNR}^{-1}_{T}\,\,\,\,\textrm{and   $\,\,\,\textrm{SNR}_{T}=\frac{NP_T}{\sigma^2_{Z}}$.}\label{eq-SNR-def}
\end{eqnarray}  

(ii) \emph{The channel affects both the training sequence and the state sequence}, which is unknown at the receiver, i.e. the decoder observes $\mb{y}_{T}=H (\mb{x}_T+\mb{s}_T)+\mb{z}_{T},$ where $\mb{s}_{T}$ is the state sequence affecting the channel as multiplicative noise. This scenario arises in robust digital watermarking where the channel means an unknown multiplicative attack on the host signal $\mb{s}_{T}$ that is used for training. Here, because the presence of $\mb{s}_T$ with average energy per symbol $Q\gg P_T$, the scenario is much complicated than (i). In other words, as a consequence of this a different method for channel estimation is needed. 

We note that the transmitter, before sending the training sequence, perfectly knows the state sequence $\mb{s}_T$. Therefore, it can be used for adapting the training sequence to reduce the multiplicative noise at the transmitter. Consider the mean estimator $\widehat{H}_\Delta = \langle \mb{y}_{T}\rangle=H \bar{\nu} + \langle \mb{z}_{T} \rangle,$ where $\bar{\nu}=\langle \mb{x}_{T}\rangle+\langle \mb{s}_{T}\rangle$ and $\langle \cdot\rangle$ denotes the mean operator. Obviously, if for some length $N$ the transmitter disposes of enough power $P_T$ to get $\bar{\nu}=1$ the interference could completely be removed from $\mb{y}_{T}$. Of course, this is not possible for all sequences $\mb{s}_T$, and only part of these sequences can be removed. We can state this more formally as the following optimization problem. Given some arbitrary pair $(\Delta,\gamma)$ with $0\leq(\Delta,\gamma)< 1$, we find the optimal training sequence $\mb{x}^*_{T}$ and its required length $N^*$ such that
 \vspace{-2mm}
\begin{eqnarray}
\left\{\begin{array}{ll} 
\textrm{Minimize}  &  \|\mb{x}_{T} \|^2/N, \\
\textrm{Subjet to} &  \Pr_{\mb{s}_{T}}  \big( \bar{\nu}^2< (1-\Delta)P_T  \big) \leq \gamma,\label{eq-estimation-2}
\end{array}\right.
\end{eqnarray}
where $(1-\Delta)P_T$ is the remainder power after removing $\mb{s}_{T}$. This means that for $100\times(1-\gamma)\%$ of estimations the interference can be removed, elsewhere the training fails. We call $\gamma$ the failure tolerance level. Then, the solution of \eqref{eq-estimation-2} is easily found to be $\mb{x}^*_{T}(\mb{s}_{T})=(x_0^*,\dots,x_0^*) $ with
\begin{equation}
x_0^*(\mb{s}_{T})=\left\{\begin{array}{ll} 
\sqrt{(1-\Delta)P_T }- \langle \mb{s}_{T}\rangle & \textrm{if $ \| \mb{x}_{T}^*(\mb{s}_{T}) \|^2\leq N P_T$}, \\
0 & \textrm{elsewise},
\end{array}\right.
\end{equation}
and $N^*$ is chosen such that $\Pr_{\mb{s}_{T}} \big( \| \mb{x}_{T}^*(\mb{s}_{T}) \|^2> N^* P_T  \big) \leq \gamma$. It follows that $N^*$ can be computed by using the cumulative function of a non-central chi-square of two degrees of freedom $\textrm{cdf}\big(r;2, 2N^*P_T(1-\Delta)Q^{-1} \big)=1-\gamma$ with $r=\frac{2N^*}{Q}P_T$. Actually, the channel estimate can be written as:
\begin{eqnarray}
\widehat{H}_\Delta &=& H+\tilde{\mc{E}}, \label{eq-error-2}
\end{eqnarray}  
where $\tilde{\mc{E}} = \big((1-\Delta)P_T\big)^{-1/2} \langle \mb{z}_{T} \rangle $ is the estimation error with
\begin{eqnarray}
\sigma_{\tilde{\mc{E}}}^2=\textrm{SNR}^{-1}_{T,\Delta}\,\,\,\,\textrm{and   $\,\,\,\textrm{SNR}_{T,\Delta}=\frac{N(1-\Delta) P_T}{\sigma^2_{Z}}$}.
\end{eqnarray}  
Note that $\sigma_{\tilde{\mc{E}}}^2=(1-\Delta)^{-1} \sigma_{\mc{E}}^2$, where $\sigma_{\mc{E}}^2$ is the estimation error in (i). To compare both estimation scenarios, we define the \emph{noise reduction factor} $\eta=\big(N(1-\Delta)\big)^{-1}$. 
\begin{figure}[thpb]
\centering
\includegraphics[width=2.8in]{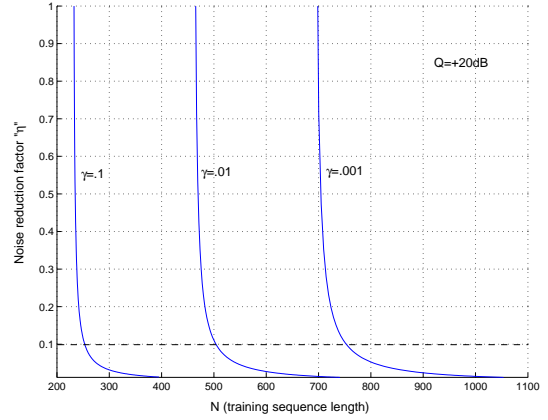}
\caption{Noise reduction factor $\eta$ vs the training sequence lengths $N$, for various probabilities $\gamma$.\vspace{-4mm}}
\label{fig_Final_training}
\end{figure}

Fig.\ \ref{fig_Final_training} shows the noise reduction factor $\eta$ versus the training sequence length $N$,  for various failure tolerance levels $\gamma\in\{10^{-1},10^{-2},10^{-3}\}$. The power of the state sequence $Q$ is $20\,$dB larger than that corresponding to the training sequence $P_T$. Let us suppose that, e.g., we want to get an estimation error $10$ times less than the channel noise (i.e. $\eta=10^{-1}$), with a failure tolerance level $\gamma=10^{-2}$. From Fig.\ \ref{fig_Final_training} we can observe that the required training length is $N=500$. Whereas in (i), where the state sequence is not present during the training, to get equal performances we would only require $N=10$. 

Finally, we characterize both channel estimation performances in terms of the {\it a posteriori} pdf of $H$ given $\widehat{H}_{\textrm{ML}}$ and the pdf of $H$ given $\widehat{H}_{\Delta}$. These pdfs will be needed in the next section to derive a composite channel model and its achievable rates. Using the fading pdf, the expression \eqref{eq-error-1} and \eqref{eq-error-2} and some algebra, we obtain
\begin{eqnarray}
\left\{\begin{array}{l} \psi_{H|\widehat{H}_{\textrm{ML}}}(H|\widehat{H}_{\textrm{ML}})=C\mc{N}(\delta\widehat{H}_{\textrm{ML}},\delta \sigma_{\mc{E}}^2),\\
\psi_{H|\widehat{H}_\Delta}(H|\widehat{H}_\Delta)=C\mc{N}(\tilde{\delta}\widehat{H}_\Delta, \tilde{\delta}\sigma_{\tilde{\mc{E}}}^2),\end{array}\right. \label{eq-pdf-aposteriori}
\end{eqnarray}  
where $\delta= (\sigma_h^2+\sigma_{\mc{E}}^2)^{-1}\sigma_h^2$ and $\tilde{\delta}=(\sigma_h^2+\sigma_{\tilde{\mc{E}}}^2)^{-1}\sigma_h^2 $. 

\section{Main results}

In this section, we first introduce the notion of reliable communication based on the average of the error probability over all channel estimation errors. This notion, for DMCs with state information non-causally known at the transmitter, allows us to consider the capacity of a composite (more noisy) channel. Then, we find the optimal DPC scheme and its achievable rates for the channel descripted in Section II with imperfect channel estimation (see Section III). 

\subsection{Problem Definition and Coding Theorem}

A message $m$ from the set $\mc{M}=\{1,\dots,\lfloor 2^{nR}\rfloor \}$ is transmitted using a length-$n$ block code defined as a pair $(\varphi,\phi)$ of mappings, where $\varphi: \mc{M}\times\mathscr{S}^n    \mapsto   \mathscr{X}^n$ is the encoder, and $\phi: \mathscr{Y}^n\times  \Theta  \mapsto   \mc{M}\cup \{0\}$ is the decoder (that utilizes $\hat{\theta}$). Note that the encoder uses the realization of the state sequence $\mathbf{s}$. This knowledge is exploited for encoding the information messages $m\in\mc{M}$. The rate, which depends on the channel estimate $\hat{\theta}$ through its decoder, is given by $n^{-1}\log_2 \Mcb_{\hat{\theta}}$. The maximum (over all messages) of the average of the error probability over all channel estimation errors
\[ \bar{e}_{\max}(\varphi,\phi,\hat{\theta})=\max_{m\in\mc{M}}\,\, \mathbb{E}_{\theta\mathbf{s}|\hat{\theta} }\big[\!\!\!\!\!\! \!\!\! \!\! \! \sum_{\mathbf{y}\in\mathscr{Y}^n:\phi(\mathbf{y}, \hat{\theta})\neq m}\!\!\!\!\!\!\!\!\!\!\!W^n\big(\mathbf{y}|\varphi(m,\mathbf{s}),\mathbf{s},\theta\big)\big].
\]

For a given channel estimate $\hat{\theta}$, and $0<\epsilon< 1$, a rate $R\geq 0$ is $\epsilon$-achievable on an estimated channel, if for every $\delta>0$ and every sufficiently large $n$ there exists a sequence of length-$n$ block codes such that the rate satisfies $n^{-1}\log M_{\hat{\theta}}\geq R-\delta$ and $\bar{e}_{\max}(\varphi,\phi,\hat{\theta}) \leq \epsilon$. This definition requires that maximum of the averaged error probability occurs with probability less than $\epsilon$. For a more robust notion of reliability over single-user channels we refer the reader to \cite{piantanida_isita2006}. Then, a rate $R\geq 0$ is achievable if it is $\epsilon$-achievable for every $0<\epsilon<1$. Let $C_{\epsilon}(\hat{\theta})$ be the largest $\epsilon$-achievable rate for a given estimated $\hat{\theta}$. The mean capacity over all channel estimates is then defined as the mean of largest achievable rate, i.e., \vspace{-1mm}
\[
\bar{C}=\lim\limits_{\epsilon\downarrow  0} \mathbb{E}_{\hat{\theta}} \big[ C_{\epsilon}(\hat{\theta})\big].
\] 
We next state a theorem quantifying this capacity.

\begin{theorem}\label{theo-capacity}
Given an estimate $\hat{\theta}$ known at the receiver and no channel information at the transmitter. The capacity of a channel $W(\cdot|x,s,\theta)$ with channel state information non-causally known at the transmitter is given by
\begin{equation}
\bar{C} =\!\!\!\!\! \max\limits_{P(u,x|s) \in \mc{P}(\mathscr{U} \! \times \! \mathscr{X})}\!\!\! \mathbb{E}_{\hat{\theta}} \big[ \mathscr{C}\big(P(u,x|s),\hat{\theta}\big)\big],\vspace{-2mm}\label{qe-capacity}
\end{equation}
where\vspace{-1mm}
\begin{equation}
 \mathscr{C}\big( P(u,x|s),\hat{\theta} \big) =I \big( P_U;\widetilde{W}(\cdot |u,\hat{\theta}) \big)-I\big(P_S; P_{U|S}\big).\label{eq-mutual-inf}
\vspace{.5cm}
\end{equation}
\end{theorem}

In this theorem $\mc{P}(\mathscr{U}\! \times \! \mathscr{X})$ denotes the set of PMs on $(\mathscr{U} \!  \times \! \mathscr{X})$ so that $U\minuso (X,S) \minuso Y$ form a Markov Chain. We emphasize that the supremum in \eqref{qe-capacity} is taken over all input distributions not depending on the channel estimates $\hat{\theta}$. The composite channel 
\begin{equation}
\widetilde{W}(y |u,\hat{\theta}) = \!\!\!\!\! \sum_{(x,s)\in\mathscr{U}\! \times \! \mathscr{X}} \!\!\!\!\! P(x|u,s)P_S(s)\widetilde{W}(y |x,s,\hat{\theta}),\label{eq-composite-model}
\end{equation}
and $\widetilde{W}(y |x,s,\hat{\theta}) = \mathbb{E}_{\theta | \hat{\theta}} \big[ W(y |x,s,\theta)  \big]$, where $\mathbb{E}_{\theta | \hat{\theta}} \big[\cdot \big] $ denotes the expectation with the conditional pdf $\psi_{\theta | \hat{\theta}}$ characterizing the channel estimation. We also used the mutual information
$$
I \big( P_U;\widetilde{W}(\cdot |u,\hat{\theta}) \big)= \sum_{u\in\mc{U}}\sum_{y\in\mc{Y}}
P(u) \widetilde{W}(y|u,\hat{\theta})\log_2 \frac{\widetilde{W}(y|u,\hat{\theta})}{ Q(y|\hat{\theta}) },
$$
with $ Q(y| \hat{\theta}) = \sum_{u\in\mc{U}} P(u)\widetilde{W}(y|u,\hat{\theta}) $. The capacity can be attained by using the maximum-likelihood (ML) decoding metric based on the composite channel model \eqref{eq-composite-model} (cf. \cite{asilomar06}). The proof of this coding theorem is straightforward from \cite{pinsker-1980} and basic information properties.

\subsection{Achievable rates and optimal DPC scheme}

We derive achievable rates for the channel \eqref{channel-model} by assuming Gaussian inputs and both estimation scenarios \eqref{eq-error-1} and \eqref{eq-error-2}. To evaluate \eqref{qe-capacity} in \eqref{channel-model} requires solving an optimization problem where we have to determine the optimum distribution $P(u,x|s)$ maximizing the capacity. We begin by computing the composite channel model for both estimation scenarios, i.e.  $\widetilde{W}(y|x,s, \widehat{H}_{\textrm{ML}}) = \mathbb{E}_{H | \widehat{H}_{\textrm{ML}}} \big[ W(y |x,s,H)  \big]$ and $\widetilde{W}(y|x,s,\widehat{H}_{\Delta})= \mathbb{E}_{H | \widehat{H}_{\Delta}} \big[ W(y |x,s,H)  \big]$. From \eqref{eq-pdf-aposteriori} it is not difficult to show that 
\begin{equation}
\widetilde{W}\big(y|x,s, \widehat{H}_{\textrm{ML}}\big) =C\mc{N}\big(\delta \widehat{H}_{\textrm{ML}}(x+s),\sigma^2_Z+\delta \sigma^2_{\mc{E}}(|x|^2+|s|^2)\big),\label{composite-channel-ML}
\end{equation}
\begin{equation}
\widetilde{W}\big(y|x,s, \widehat{H}_{\Delta}\big) =C\mc{N}\big(\tilde{\delta} \widehat{H}_{\Delta}(x+s),\sigma^2_Z+\tilde{\delta} \sigma^2_{\tilde{\mc{E}}}(|x|^2+|s|^2)\big).\label{composite-channel-DELTA}
\end{equation}
Actually, we only need to consider the capacity associated to \eqref{composite-channel-ML} corresponding to the scenario (i), since the pdf \eqref{composite-channel-DELTA} differences in constant quantities. 

\emph{Channel estimates known at the transmitter:} Obviously, if the channel estimates $\widehat{H}_{\textrm{ML}}$ are known at the transmitter, the optimal input distribution is shown to be given by
\begin{equation}
P_{\widehat{H}_{\textrm{ML}}}\big(u,x|s\big)=\left\{\begin{array}{ll} P(x)  & \textrm{if $\,\,\,u=x+\alpha^*( \widehat{H}_{\textrm{ML}}) s$}, \\
0 & \textrm{elsewhere},
\end{array}\right.\label{optimal-DPC-TxRx}
\end{equation}
where $P(x)=C\mc{N}\big(0,\bar{P}\big)$, and $\bar{P}$ is the power constraint and
\begin{equation}
\alpha^*( \widehat{H}_{\textrm{ML}})=\frac{\delta^2 |\widehat{H}_{\textrm{ML}}|^2 \bar{P}}{\delta^2 |\widehat{H}_{\textrm{ML}}|^2 \bar{P}+ \sigma^2_Z+\delta \sigma^2_{\mc{E}}(\bar{P}+Q)}.\label{optimal-alpha-DPC-TxRx}
\end{equation}
The capacity denoted $\bar{C}_{TxRx}$ is then
\begin{equation}
\bar{C}_{TxRx}=\mathbb{E}_{\widehat{H}_{\textrm{ML}}} \Big\{ \log_2\left(1+ \frac{\delta^2 |\widehat{H}_{\textrm{ML}}|^2 }{ \sigma^2_Z+\delta \sigma^2_{\mc{E}}(\bar{P}+Q)}\right) \Big\}.\label{eq-Ctxrx}
\end{equation}
This easily follows from the fact that in this case it is possible to swap expectation and maximization in \eqref{qe-capacity}.

\emph{Channel estimates unknown at the transmitter:} Here we cannot use the optimal DPC scheme \eqref{optimal-DPC-TxRx}, because the channel estimates $\widehat{H}_{\textrm{ML}}$ are not available at the transmitter to compute the parameter \eqref{optimal-alpha-DPC-TxRx}. However, assuming Gaussian inputs, which means that $P\big(u,x|s\big)$ is a conditional joint Gaussian pdf. The optimal DPC scheme can be shown to be given by
\begin{equation}
P\big(u,x|s\big)=\left\{\begin{array}{ll} P(x)  & \textrm{if $\,\,\,u=x+\alpha s$}, \\
0 & \textrm{elsewhere},
\end{array}\right. \label{optimal-DPC-Rx}
\end{equation}
where $\alpha\in[0,1]$ is the parameter maximizing \eqref{qe-capacity}. Hence, given $\alpha$ the achievable rates can be computed by replacing \eqref{composite-channel-ML} and \eqref{optimal-DPC-Rx} in \eqref{eq-mutual-inf}. Thus, using some algebra we obtain 
\begin{equation}
I_{\alpha} \big( P_U;\widetilde{W}(\cdot |u,\hat{\theta}) \big)=\log_2\left( \frac{(\mathbb{P}+\mathbb{Q}+\mathbb{N})(\mathbb{P}+\alpha^2 \mathbb{Q})}{\mathbb{P}\mathbb{Q}(1-\alpha)^2+\mathbb{N}(\mathbb{P}+\alpha^2 \mathbb{Q})} \right),\label{eq-inf-1}
\end{equation}
\begin{equation}
I_{\alpha} \big(P_S; P_{U|S}\big)=\log_2\left( \frac{\mathbb{P}+ \alpha^2 \mathbb{Q}}{\mathbb{P}}\right),\label{eq-inf-2}
\end{equation}
where $\mathbb{P}=\delta^2 |\widehat{H}_{\textrm{ML}}|^2 \bar{P}$,  $\mathbb{Q}=\delta^2 |\widehat{H}_{\textrm{ML}}|^2 Q$ and $\mathbb{N}= \sigma^2_Z+\delta \sigma^2_{\mc{E}}(\bar{P}+Q)$. Given $0\leq\alpha\leq 1$, by using \eqref{eq-inf-1} and \eqref{eq-inf-2}, the capacity $\bar{C}_{Rx}(\alpha)$ writes
\begin{equation}
\bar{C}_{Rx}(\alpha)=\mathbb{E}_{\widehat{H}_{\textrm{ML}}} \Big\{ \log_2\left(\frac{\mathbb{P}(\mathbb{P}+\mathbb{Q}+\mathbb{N})}{\mathbb{P}\mathbb{Q}(1-\alpha)^2+\mathbb{N}(\mathbb{P}+\alpha^2 \mathbb{Q})}\right) \Big\}. \label{alpha_capacity}
\end{equation}
We remark that our Gaussian assumption only leads to a lower bound \eqref{alpha_capacity} of the capacity \eqref{eq-composite-model}. However, in the next section we shall observe that this bound is tight for realistic SNR values. Actually, it remains to find the optimal parameter $\alpha$ maximizing \eqref{alpha_capacity}. Let us first consider the more intuitive suboptimal choice given by the mean of the optimal $\alpha^*( \widehat{H}_{\textrm{ML}})$ in \eqref{optimal-alpha-DPC-TxRx}, i.e. $\bar{\alpha}=\mathbb{E}_{\widehat{H}_{\textrm{ML}}} \big\{ \alpha^*( \widehat{H}_{\textrm{ML}}) \big\}$. To compute this mean, we note that $ \widehat{H}_{\textrm{ML}}$ has a Gaussian pdf $C\mc{N}\big(0, \sigma_h^2+\sigma^2_{\mc{E}}\big)$. Hence, we can show that\vspace{-2mm}
\begin{equation}
\begin{array}{ll}
\bar{\alpha}=1- \rho \exp (\rho) E_1 (\rho), & \textrm{ with $\rho=\displaystyle{\frac{\mathbb{N}}{\delta^2 \bar{P} (\sigma_h^2+\sigma^2_{\mc{E}})}}$}, \label{mean-alpha}
\end{array}
\end{equation}
where $E_1(z)=\int\limits_z^\infty t^{-1}\exp(-t) dt$ denotes the exponential integral function. Therefore, all rates smaller than $\bar{C}_{Rx}(\bar{\alpha})$ are achievable by using DPC scheme \eqref{optimal-DPC-Rx} and the mean $\bar{\alpha}$ \eqref{mean-alpha}.  

Another possibility is to find directly the optimal parameter $\alpha^*$ maximizing \eqref{alpha_capacity}. To this end, we observe that
\begin{equation}
\alpha^*=\arg\min\limits_{0 \leq \alpha \leq 1} \mathbb{E}_{\widehat{H}_{\textrm{ML}}} \Big\{ \log_2 \left( \mathbb{P}\mathbb{Q}(1-\alpha)^2+\mathbb{N}(\mathbb{P}+\alpha^2 \mathbb{Q}) \right) \Big \}.\label{eq-aux}
\end{equation}
Using some algebra, from \eqref{eq-aux}, we can obtain
\begin{equation}
\begin{array}{ll}
\alpha^*=\arg\min\limits_{0 \leq \alpha \leq 1} \Big\{ \log_2 (\bar{P}/Q+\alpha^2)+\\ 
 \displaystyle{\frac{1}{\log(2)}\exp\left(\frac{\rho(\bar{P}/Q+\alpha^2)}{(1-\alpha)^2} \right)E_1 \left(\frac{\rho(\bar{P}/Q+\alpha^2)}{(1-\alpha)^2} \right)} \Big \}.\label{optimal-alpha}
\end{array}
\end{equation}
Unfortunately, there is not exists an explicit solution for $\alpha^*$ in \eqref{optimal-alpha}. However, this maximization can be numerically solved to then compute $\bar{C}_{Rx}(\alpha^*)$.

All derived results through this section are also valid for the channel model \eqref{composite-channel-DELTA}, corresponding to the estimation scenario (ii). We replace $\delta$ with $\tilde{\delta}$ and $\sigma^2_{\mc{E}}$ with $\sigma^2_{\tilde{\mc{E}}}$ in all expressions. 
\begin{figure}[thpb]
\centering
\includegraphics[width=3in]{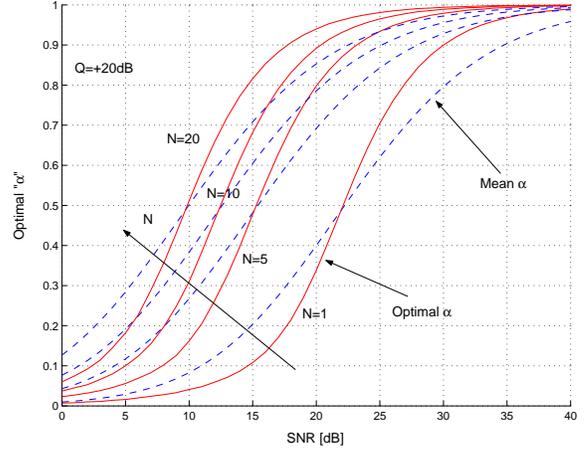}
\caption{Optimal parameter $\alpha^*$ (solid lines) vs the SNR, for various training sequence lengths $N$. Dashed lines show mean alpha $\bar{\alpha}$.\vspace{-4mm}}
\label{fig_alpha}
\end{figure}

\section{Simulation results and discussions}

In this subsection, numerical results are presented based on Monte Carlo simulations. 
Fig. \ref{fig_alpha} shows both the mean parameter $\bar{\alpha}$ \eqref{mean-alpha} and the optimal parameter $\alpha^*$ \eqref{optimal-alpha} versus the signal-to-noise ratio, for various training sequence lengths $N$. The state sequence power $Q$ is $+20\,$dB larger than that of the channel input $\bar{P}$, and the training power is $P_T=\bar{P}$. We can observe that both parameters are relatively close for many SNR values. Furthermore, even in the SNR ranges where the values seem to be quite different, we have observed that the achievable rates with $\bar{\alpha}$ are very close to those provided by the optimal solution $\alpha^*$. Therefore, we can conclude that the mean parameter can be used to design the optimal DPC scheme. 
 
\begin{figure}[thpb]
\centering
\includegraphics[width=3in]{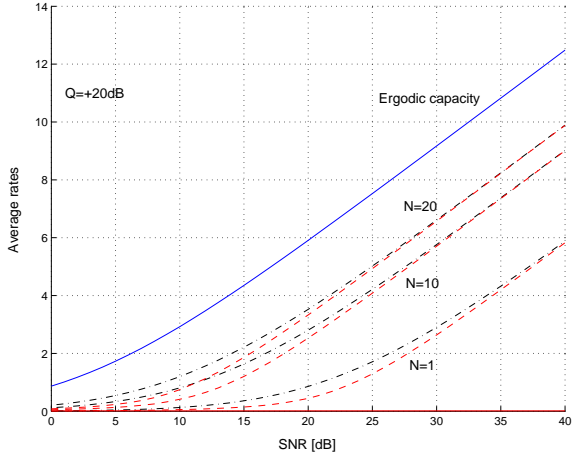}
\vspace{-2mm}\caption{Achievable rates with channel estimates known at the transmitter (dashed-dot lines) vs the SNR, for various training sequence lengths $N$. Dashed lines suppose channel estimates unknown at the transmitter. Solid line shows the capacity with the channel known at both transmitter/receiver.\vspace{-6mm}}
\label{fig_Final_capacity}
\end{figure}

Fig. \ref{fig_Final_capacity} shows achievable rates \eqref{alpha_capacity} (in bits per channel use) with channel estimates unknown at the transmitter versus the SNR, for various training sequence lengths $N\in\{1,10,20\}$ (dashed line). For comparison we also show achievable rates \eqref{eq-Ctxrx} with channel estimates known at the transmitter (danshed-dot line) and with perfect channel knowledge at both transmitter and receiver (solid line). It is seen that the average rates tend to increase rather fast with the amount of training. For example, to achieve $2$ bits with channel estimates unknown at the transmitter. Observe that a scheme with estimated channel and $N=10$ requires $18\,$dB, i.e., $11\,$dB more than with perfect channel information. Whereas, if the training length is further reduced to $N=1$, this gap increases to $27$\,dB. On the other hand, when the channel estimates are known at the transmitter, the SNR requeried for $2$ bits is only $1\,$dB less than the case with channel estimates unknown. This rate gain is slightly smaller, and consequently we can conclude that the knowledge of the channel estimates at the transmitter is not really necessary with the proposed DPC scheme. 

\begin{figure}[thpb]
\centering
\includegraphics[width=3in]{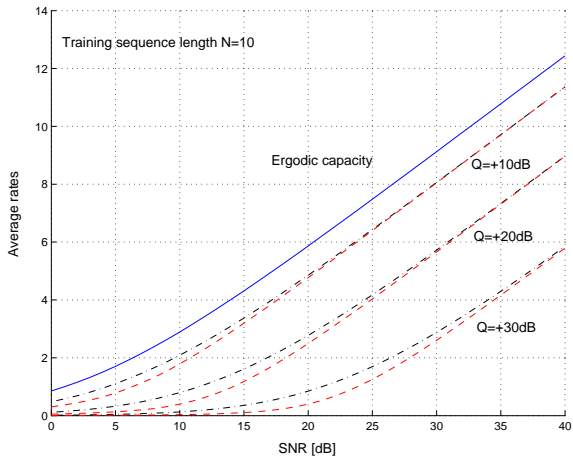}
\vspace{-2mm}\caption{Similar plots for different power values of the state sequence $Q$.\vspace{-10mm}}
\label{fig_Final_comp_Q_cap}
\end{figure}

Finally, we study the impact of the power state sequence on the achievable rates. Fig. \ref{fig_Final_comp_Q_cap} shows similar plots for different values of $+Q\in\{+20,+30,+40\}$, i.e., $Q$ is times larger (in dB) than the channel input power $P$, and training sequence length is $N=10$. We can observe that the performance are very sensitive to the power $Q$. This is because with imperfect channel estimation the capacity still depends on $Q$ (cf. \eqref{alpha_capacity}), while with perfect channel information the state sequence is cancelled at the transmitter independent of the power $Q$. 
\vspace{-2mm}

\vspace{-3mm}
\section{Conclusion}

In this paper we studied the problem of communicating reliably over unknown channels with channel states non-causally known at the transmitter. We assumed that no channel information is available at the transmitter and imperfect channel information is available at the receiver, i.e., the receiver only has access to a noisy estimate of the channel. In this scenario, we proposed to characterize the information-theoretic limits through the notion of reliable communication based on the average of the error probability over all channel estimation errors. We presented an explicit expression, for general DMCs, of its maximal achievable rate averaged over all channel estimates. Then, we computed mean achievable rates for the fading Costa's channel with ML channel estimation and Gaussian inputs. We also studied optimal training design adapted to each application, e.g. Broadcast Channels or watermarking. 

The somewhat unexpected result is that, while it is well-known that DPC requires perfect channel knowledge at both transmitter and receiver, without channel information at the transmitter, significant gains can be still achieved by the DPC strategy, using the proposed DPC scheme. Further numerical results show that, under the assumption of imperfect channel information at the receiver, the benefit of channel estimates known at the transmitter does not lead to large rate increases. 

Codes achieving capacity do not need to be long to exploit the long-term ergodic properties though the estimated fading process, and can be applied when the real transmission time is not large compared to the coherence time of the channel.


\bibliographystyle{ieeetr}

\bibliography{biblio}

\end{document}